\documentclass[sigconf,nonacm]{acmart}
\usepackage{enumitem}
\usepackage{etoolbox}
\providecommand{\aptLtoX}[3][]{#3}

\setcopyright{none}
\makeatletter
\let\@copyrightspace\relax
\makeatother

\apptocmd{\maketitle}{%
  \footnotetext{%
  © Mattias Rost 2026. This is the author's version of the work. The definitive version will be published in \emph{Proceedings of the 2026 CHI Conference on Human Factors in Computing Systems (CHI~'26)}, Association for Computing Machinery (ACM), https://doi.org/10.1145/3772318.3791769.
  }%
}{}{}

\begin{document}

\title{Co-Disclosing the Computer: LLM-Mediated Computing through Reflective Conversation}

\author{Mattias Rost}
\email{mattias.rost@ait.gu.se}
\orcid{0000-0002-6842-0226}
\affiliation{%
  \institution{University of Gothenburg}
  \city{Gothenburg}
  \country{Sweden}
}

\begin{abstract}
Large language models (LLMs) are changing how we interact with computers. As they become capable of generating software dynamically, they invite a fundamental rethinking of the computer's role in human activity. In this conceptual paper, we introduce LLM-mediated computing: a paradigm in which interaction is no longer structured around fixed applications, but emerges in real-time through human intent and LLM interpretation. We make three contributions: (1) we articulate a new interaction metaphor of reflective conversation to guide future design, (2) we use the lens of postphenomenology to understand the human-LLM-computer relation, and (3) we propose a new mode of computing based on co-disclosure, in which the computer is constituted in use. Together, they define a new mode of computing, provide a lens to analyze it, and offer a metaphor to design with.
\end{abstract}



\maketitle

\section{Introduction}

\textit{What is the computer becoming when its behavior is generated in real time in response to human intent?}

Recent advances in large language models (LLMs) have made it possible to generate not only text, but also code, interface structures, and software behavior in response to natural language prompts \cite{chenEvaluatingLargeLanguage2021a,ouyangTrainingLanguageModels2022}. While many systems currently situate LLMs inside applications or use them as chat-based agents such as ChatGPT, we propose a broader shift: one in which the LLM functions as a mediator between human intent and emergent computational form. We call this \textit{LLM-mediated computing} \cite{rostReclaimingComputerLLMMediated2025}. In this paradigm, the computer is no longer a pre-configured environment of tools, but a relational system whose structure and affordances are constituted through use. Computing becomes situated, dialogic, and interpretive. Functionality is not invoked in advance, but is brought forth in response to ongoing activity. Read hermeneutically, this reframes interaction from operating fixed representations to a process of co-interpretation in which meaning and capability are disclosed together over time. Like a conversation, each move projects relevant next moves, setting up a sequential structure of responsiveness and repair rather than one-off execution.

To understand this shift, we must reconsider one of the most enduring metaphors in human-computer interaction: \textit{the desktop} \cite{johnsonXeroxStarRetrospective1989}. For over half a century, application-centered computing has structured interaction around stable interfaces such as word processors, browsers, and spreadsheets, each representing a fixed slice of capability. This model has enabled learnability and predictability, but it has also constrained how we imagine what computing can be. It fragments user activity across tools and positions the computer as a static substrate that is acted upon.

LLM-mediated computing challenges this paradigm. It reframes the computer not as a container of applications, but as a relational configuration that emerges through a triadic structure: human activity, LLM interpretation and code generation, and the resulting transformation of the computational environment. We define LLM-mediated computing as a mode of interaction in which large language models dynamically generate both executable code and interface structures, transforming the computer’s capabilities in real time as users express intent.

This shift has profound implications for how we think about technology, design systems, and relate to digital environments. To make sense of it, we bring together insights from the philosophy of technology, especially postphenomenology and the phenomenology of intentionality, with critical perspectives from HCI. We argue that understanding this shift requires more than technical analysis: it calls for conceptual and philosophical reflection.

In this conceptual paper, we make three contributions:


\aptLtoX[graphic=no,type=html]{
\begin{enumerate} 
\item[] \makebox[2em][l]{\textbf{C1}} A new interaction metaphor: \textit{reflective conversation}. Drawing on Donald Schön \cite{schonReflectivePractitionerHow2017}, we propose this metaphor as an alternative to the desktop, one that foregrounds responsiveness, temporality, and shared activity.
\item[] \makebox[2em][l]{\textbf{C2}} A postphenomenological analysis of the human–LLM–computer relation, extending theories of technological mediation to account for generative and interpretive systems.
\item[] \makebox[2em][l]{\textbf{C3}} A new mode of computing based on \textit{co-disclosure}. We argue that the computer, when mediated by an LLM, is not pre-given but co-constituted in use.
\end{enumerate}
}{
\begin{enumerate}[label={C\arabic*}]
  \item
  A new interaction metaphor: \textit{reflective conversation}. Drawing on Donald Schön \cite{schonReflectivePractitionerHow2017}, we propose this metaphor as an alternative to the desktop, one that foregrounds responsiveness, temporality, and shared activity.
  
  \item
  A postphenomenological analysis of the human–LLM–computer relation, extending theories of technological mediation to account for generative and interpretive systems.
  
  \item
  A new mode of computing based on \textit{co-disclosure}. We argue that the computer, when mediated by an LLM, is not pre-given but co-constituted in use.
\end{enumerate}
}

Together, these contributions lay a foundation for understanding and designing in an era where the boundaries of the computer are no longer drawn by applications, but emerge through interpretive interaction. They move us beyond application-centered paradigms and toward a model of computing that is more fluid, relational, and aligned with human intentionality.

Alongside these conceptual contributions, we also preview the ethical and political stakes of this shift: questions of responsibility, the opacity of mediation, and the politics of whose norms and voices are embedded in LLM-mediated computing.

\section{Background \& Related Work}

In order to understand LLM-mediated computing, we must first describe where it is coming from. Our proposal of LLM-mediated computing as co-disclosure builds on, but also departs from, several adjacent traditions.

\subsection{Computers, Software, and the Desktop Metaphor}

Computing shifted from hard-wired machines to stored-program architectures in the mid-20th century, making behavior reconfigurable through code rather than rewiring hardware \cite{vonneumannFirstDraftReport1982}. As personal computing spread, graphical user interfaces and the desktop metaphor (files, folders, applications) made systems accessible to non-experts \cite{johnsonXeroxStarRetrospective1989,smithPygmalionCreativeProgramming1975}. While this move democratized computing, it also followed an application-centered logic: capabilities are packaged in advance, interfaces present fixed affordances, and users adapt intentions to available tools. It reinforced the idea that interaction consists of manipulating predefined representations rather than exploring the machine as an open-ended medium.

This representational logic has been critiqued extensively within HCI. Winograd and Flores \cite{winogradUnderstandingComputersCognition1986} argued that it does not align with how human activity unfolds in the world. Drawing on phenomenology, they proposed that meaning emerges through engagement in action rather than through abstract manipulation of internal models. Dourish extended this critique by introducing the concept of embodied interaction, suggesting that interfaces should support participation in ongoing activity rather than observation of symbolic systems \cite{dourishWhereActionFoundations2001}. Suchman similarly showed how plans are not causes of action but resources for interpretation, highlighting the improvisational nature of real-world practice \cite{suchmanHumanMachineReconfigurationsPlans2007}. From a different perspective, Bardzell and Bardzell have argued that representational approaches also encode normative assumptions about users and tasks \cite{bardzellHumanisticHCI2015}. They call for interaction design that embraces ambiguity, multiplicity, and cultural specificity rather than treating users as generic agents.

The persistence of representational systems points to a disconnect between theory and practice. As Harrison, Tatar, and Sengers describe, HCI has moved through multiple paradigms, from optimizing fit between humans and machines, to modeling cognition, to understanding activity as embodied and situated \cite{harrisonThreeParadigmsHCI2007}. Yet despite these shifts, most interfaces continue to reflect the assumptions of the second paradigm. They rely on fixed metaphors and predefined structures that limit how computing can be understood and experienced. In this paper, we suggest that LLM-mediated computing invites a re-examination of these assumptions. Rather than relying on static representations, it enables interaction to unfold through interpretation and responsiveness. In doing so, it moves closer to the commitments of the third paradigm by supporting relational and emergent forms of engagement.

More recent work on Entanglement HCI has even proposed this orientation as a possible “fourth paradigm”, emphasizing that humans and technologies are inseparably bound in sociomaterial assemblages rather than separated into users and tools \cite{frauenbergerEntanglementHCINext2019,bardzellHumanisticHCI2015}. This perspective strengthens the critique of representational metaphors by showing how interaction is always already embedded in webs of relations. Our own proposal resonates with this shift, but as we argue later, LLM-mediated computing extends the notion of entanglement to the “computer” itself, where the very functionality of the computer is co-disclosed in temporally unfolding interaction.


From a postphenomenological perspective, the desktop metaphor installs a hermeneutic relation (see \cite{ihdeTechnologyLifeworldGarden1990}) as the default mode of computing: users read icons, menus, and files as signs that stand in for underlying operations. This stabilizes the computer as a representational text to be interpreted rather than a medium to be reconfigured. The very success of the desktop thus established representation as the dominant way the computer appears.

Such stable representations are, however, not necessarily stable in use but appropriated \cite{carrollCompletingDesignUse2004}. Studies of appropriation in HCI have shown that users often reinterpret and adapt technologies beyond designer‐intended purposes. For example, Salovaara \cite{salovaaraAcceptanceAppropriationDesignOriented2009,salovaaraEverydayAppropriationsInformation2011} argues that appropriation involves creative use and cognitive processes of problem-solving and perception, rather than passive acceptance of artefacts. Likewise, Krischkowsky et al. \cite{krischkowskyCaptologyTechnologyAppropriation2016} emphasises the importance of unanticipated users, uses, and circumstances, and argues for designing systems that support or at least tolerate emergent appropriation. This suggests that while the desktop metaphor frames interaction as representational, actual use often involves more fluid and situated engagements. LLM-mediated computing builds on this insight by making reconfiguration an explicit part of the interaction loop, rather than an occasional appropriation.

\subsection{Historical Accounts of the Computer as a Medium}

HCI has long explored how computing environments might be made more malleable. Beaudouin-Lafon’s notion of instrumental interaction emphasized that interfaces should support the creation of new instruments in use, rather than fixing functionality in advance \cite{beaudouin-lafonInstrumentalInteractionInteraction2000}. HCI researchers have similarly argued for malleable software substrates that users can reconfigure to suit their own practices, while end-user development research has sought to empower users to tailor systems to their needs \cite{liebermanEndUserDevelopment2006,koStateArtEnduser2011}. These approaches anticipate our concern with moving beyond static applications, yet they typically rely on explicit user manipulation or authoring.

Another adjacent tradition is the vision of the computer as an open medium for construction. Alan Kay’s Dynabook concept and Papert’s constructionism framed computing as a space where learners and designers could build their own representations and simulations \cite{kayEarlyHistorySmalltalk1993,papertMindstormsChildrenComputers2020}. Direct manipulation interfaces \cite{shneidermanDirectManipulationStep1983} and later live programming environments \cite{mcdirmidLivingItLive2007,chughProgrammaticDirectManipulation2016} extended this vision by enabling users to see the effects of code and design choices immediately. Similarly, programming-by-example systems such as Pygmalion \cite{smithPygmalionCreativeProgramming1975} and Watch What I Do \cite{cypherWatchWhatProgramming1993} aimed to make computational authoring more situated and accessible. These traditions position the computer as a medium for live construction, but functionality still depends on the user explicitly creating or demonstrating behavior. 

This lineage culminates in the influential work of Bret Victor, whose essays and prototypes such as Magic Ink and Learnable Programming \cite{victorMagicInk2006,victorLearnableProgramming2012} reframed the computer as a live, epistemic medium. Unlike earlier traditions that emphasized giving users the ability to explicitly reconfigure systems, Victor argued that feedback, liveness, and immediacy should be built into the medium itself as conditions for reflective practice. LLM-mediated computing extends this trajectory: generativity and interpretive mediation become integral properties of the computer, so that reconfiguration is not an option enabled by design but an ordinary condition of interaction.


\subsection{Related Perspectives and Differentiation}

Prior work in HCI has engaged with similar questions of technological mediation, configurability, and reflection. Postphenomenology has been applied not only as a conceptual lens (e.g., \cite{ihdeTechnologyLifeworldGarden1990,verbeekWhatThingsPhilosophical2005}) but also in empirical analyses of interactive systems \cite{kudinaEthicsGoogleGlass2019}. Similarly, Reflective Design \cite{sengersReflectiveDesign2005a}, situated within broader Critical Design traditions \cite{dunneSpeculativeEverythingDesign2013}, has advocated for systems that make their assumptions visible. Yet in LLM-mediated computing, interpretive mediation arises spontaneously in the flow of action. This resonates with broader third-paradigm perspectives in HCI, which emphasize situated and embodied activity over representational models \cite{dourishWhereActionFoundations2001,kuuttiTurnPracticeHCI2014,suchmanHumanMachineReconfigurationsPlans2007}. Our account extends these by specifying how such situatedness is operationalized through LLM-driven interpret–generate–reconfigure loops that reconfigure the computer itself. This convergence of configurability, mediation, and reflection suggests that LLM-mediated computing represents a novel synthesis of prior trajectories, while introducing distinctive challenges of responsibility, legibility, and accountability.

Beyond phenomenology and reflective design, post-WIMP research traditions such as reality-based interaction \cite{jacobRealitybasedInteractionFramework2008} and activity-centered design \cite{bodkerWhenSecondWave2006}, likewise argued that computing should be grounded in lived practice. These frameworks sought to close the gap between human experience and computational substrates, often through tangible, embodied, or practice-centered approaches. Classical HCI work on mixed-initiative systems and end-user programming explored how systems can share initiative or be tailored by users  \cite{fischerMetadesignFrameworkFuture2006,horvitzPrinciplesMixedinitiativeUser1999,nardiSmallMatterProgramming1993}. Conversational interfaces and CA-informed HCI research examined turn-taking, repair, and grounding in human–system exchanges \cite{amershiGuidelinesHumanAIInteraction2019,porcheronVoiceInterfacesEveryday2018}. Together, these lines of work have emphasized configurability, shared initiative, and situated interaction as central concerns in HCI.

Contemporary AI scholarship underscores concerns with uncertainty, provenance, and contestability \cite{benderDangersStochasticParrots2021,liaoQuestioningAIInforming2020,poursabzi-sangdehManipulatingMeasuringModel2021}. Work on prompt-as-programming and agentic LLMs demonstrates the potential of dynamic orchestration \cite{liangPromptsArePrograms2025,schickToolformerLanguageModels2023}, while critical HCI highlights the infrastructural politics of data and design \cite{birhaneAlgorithmicInjusticeRelational2021,crawfordAtlasAIPower2021}. Our contribution is to bring these insights inside the interaction loop: treating provenance, repair, and contestability not as optional features but as structural conditions for reflective conversation. In this way, we reframe LLM-mediated interaction as a new paradigm where the computer itself is co-disclosed, sequentially and provisionally, in use.

This development aligns with Nielsen's observations that LLMs inaugurate a new interaction paradigm beyond direct manipulation \cite{nielsenAIFirstNew}. Yet while Nielsen frames this shift primarily in terms of interface modality, we propose that it represents a deeper reconfiguration of computing itself: from predefined applications to emergent, interpretively mediated environments.

\section{LLM-Mediated Computing: Concept and Examples}

Having outlined where LLM-mediated computing departs from application-centered models, we now turn to what it looks like in practice. In this section we define the concept more precisely and illustrate it through examples that show how computing environments can be dynamically constructed in use.

LLM-mediated computing is enabled by two capacities of large language models that have rapidly matured in recent years: (1) the ability to interpret human intent within rich contexts \cite{ouyangTrainingLanguageModels2022}, and (2) the ability to generate functional code across a wide range of programming languages \cite{chenEvaluatingLargeLanguage2021a}. These have led to claims of leading to “the end of programming” \cite{jonssonEndProgrammingWe2025,welshEndProgramming2023}. To us, they enable LLM-mediated computing.

\subsection{What Is LLM-Mediated Computing?}

LLM-mediated computing refers to a mode of interaction in which an LLM dynamically generates both executable code and user interface structures in direct response to a user’s expressed intent. In this paradigm, the LLM does not merely suggest commands or retrieve information. It acts as an interpretive engine that materializes functionality and continuously reconfigures the computing environment itself.
It treats the LLM as an active medium through which the computer’s capabilities are constituted in real time. In other words, software becomes an emergent product of interaction. The LLM generates code in response to user input, the computer executes that code immediately, and the LLM can continue to modify or extend the environment as the interaction unfolds.


This capacity is not limited to textual prompting. The LLM can interpret gestures, drawing, voice, or patterns of use as signals of intent and respond by reconfiguring the environment in real time. Each action by the user, whether explicit or implicit, can become the basis for further system transformation.

At its core, this approach relies on a generative feedback loop:

\begin{itemize}
\item The user articulates an intention through language, sketching, or other input.
\item The LLM interprets this and produces executable code or markup.
\item The computer executes the code, reconfiguring the environment.
\item The user engages with the new structures, with each interaction becoming further input for interpretation.
\end{itemize}
Through this loop, the computer becomes a medium of construction rather than a static container of applications. Functionality is not merely retrieved but assembled dynamically in context.


\subsection{Examples: Dynamic Construction of Computing Environments}

In LLM-mediated computing, functionality is not retrieved from a fixed set of tools but assembled on demand. The computer-LLM configuration becomes a responsive substrate, reconfigurable in real time to accommodate evolving activity.

\subsubsection{Example 1: Composing an Email from Context}




As a simple, near-term example, consider the following:

\begin{quote}
\textit{A user opens a blank workspace on their laptop. They type in the first line: ``To Stephanie.'' The system recognizes this as communicative intent and instantiates an email header at the top of the workspace, automatically filling in Stephanie's email address from the user's contacts. The original text is relabeled as the subject line, and a message body field appears beneath it. The user continues typing in the body: ``Remember to stop by the shop on your way home.'' Drawing on recent activity, the system recalls that the couple share a shopping list. Just below the message body, a subtle panel appears: ``You and Stephanie share a shopping list from yesterday. Add it?'' with buttons for ``Preview list'' and ``Insert into email''. When the user clicks ``Insert'', the list is pasted into the email body and a short summary line is added at the top. The user then writes: ``If anything is missing, just add it to the list and I will see it''.}

\end{quote}

At the start of this interaction there is no email client in the conventional sense, only a blank environment. Through engagement, the capability to send an email is instantiated, enriched with list sharing, and made configurable. The LLM interprets the user's free-form text, generates the email-specific structures (header, body, list integration, live-linking behavior), and the computer is reconfigured accordingly.

\subsubsection{Example 2: The Garden Vignette}

As a second example, consider the following vignette, that we refer to as the garden vignette and that we will come back to throughout the paper:


\begin{quote}
\textit{A user is sketching a garden layout on a tablet. They draw irregular shapes with a stylus: paths, flowerbeds, trees, and water features. As they draw, the system begins to interpret the sketch in context. Soft suggestions appear near the drawing: icons for plant types, options to simulate seasonal growth, and annotations offering spatial recommendations. A slider labeled ``Season'' fades in at the edge of the canvas, allowing the user to visualize changes from spring to fall.}

\textit{When the user draws a rough oval for a pond, animated ripples are added and a small note appears: ``This depth may require filtration. Add water filtration?'' Tapping the note inserts a filtration symbol and opens a side panel. Later, when the user roughly sketches a human figure next to the layout and writes ``Show to Maria'', the system offers to share the sketch, prompting a list of likely collaborators and an option to attach a short message.}
\end{quote}


In this vignette, the user's gestures and partial sketches are interpreted not merely as static input but as evolving expressions of intent. The system's responses reconfigure the environment: what begins as a blank drawing surface becomes a simulation tool, then a layout assistant, then a communication channel. At each step, the LLM interprets strokes, labels, and context, generates interface elements and behaviors (e.g., the seasonal slider, filtration module, sharing prompt), and the computer is reconstituted around the unfolding activity.

While the email scenario demonstrates dynamic construction in a familiar communicative context, the garden vignette illustrates how such dynamics can extend to more complex, multimodal activities of design and simulation. These examples illustrate a key departure from prior paradigms: the system’s capabilities and interface are not predefined objects but emergent constructions. The boundaries of what the computer is – its functions, representations, and affordances – are negotiated in real time. This is not simply a more flexible form of software. It is a fundamentally different mode of computing, in which the system itself becomes a provisional, co-constructed medium.

\subsection{Human-LLM-Computer Triad}

LLM-mediated computing reorganizes the traditional dyad of human-computer interaction into a triadic structure consisting of the human, the LLM, and the emergent computer. These are not three independent components, but interdependent participants in a configuration that gives shape to interaction. The human expresses intent through language, gesture, or action. The LLM interprets that intent and produces code, interface elements, or other generative responses. The computer environment is reconfigured accordingly, becoming the site where interaction unfolds.

This triadic relation differs from the classical dyad in two important ways. First, interaction is no longer with a computer that allows manipulation of objects the computer represents. Instead, the computer itself is subject to ongoing configuration. Second, the LLM functions not as a tool or end-user application, but as an interpreter that continuously reshapes the environment in response to human activity.

The need for this triadic structure arises because the classical dyad presumes the computer to be a stable substrate that mediates human–world relations. In LLM-mediated computing, however, the computer itself is mutable: what counts as “the computer” changes as it is continuously reconfigured. The LLM cannot simply be collapsed into the computer, since its role is interpretive rather than environmental: it translates human intent into new configurations that become the computer in that moment. Likewise, it cannot be collapsed into the human, since its generative interpretations introduce initiatives not fully reducible to user input. The triad is therefore required to make visible how human, LLM, and computer are co-constituted in interaction.

Although the relation is triadic, specific analyses sometimes foreground a pairing within the triad, for example the interpret--generate--execute configuration between the LLM and the computer. These pairings are analytic lenses rather than alternative models.

In the sections that follow, we analyse this reconfiguration of computing through a postphenomenological lens and propose the metaphor of \textit{reflective conversation} as a way to understand this new paradigm of human-computer interaction.

\section{From Conversation to Reflective Conversation}

Reflective conversation captures how LLM-mediated computing unfolds: as a temporally sequenced, reciprocal exchange in which user intent and system configuration evolve together. Where the desktop metaphor presents a static repertoire to be read and operated, reflective conversation emphasizes negotiation, responsiveness, and co-disclosure. It helps us to see computing not as operating a toolset but as participating in an unfolding exchange where the computer itself is constituted in use. We begin by describing how conversation structures interaction, then extend this to the metaphor of reflective conversation.

\subsection{Toward a Relational and Temporal Metaphor}

We use conversation not as a UI style but as an interactional structure with well-described properties. Classic conversation analysis shows that interaction unfolds through sequential organization rather than isolated commands: turns display what they are doing and make next actions conditionally relevant (e.g., question→answer) \cite{sacksSimpleSystematicOrganisation1974}. In LLM-mediated computing, this structure appears as:

\begin{itemize}
\item \textit{Turn-taking \& projectability}. Each user act (utterance, sketch, selection) projects what counts as a relevant next move; the system’s response should be projectable from that context rather than a-contextual completion.
\item \textit{Adjacency pairs}. Many exchanges form first-pair / second-pair parts (request→offer; problem→proposal). Interfaces can display the expected second pair (e.g., show “simulate”, “share”, or “revise” after a design move) to make next-step relevance visible.
\item \textit{Repair}. Breakdowns are handled through self- and other-initiated repair (clarify, restate, constrain) \cite{schegloffPreferenceSelfCorrectionOrganization1977}. Systems should surface repair moves (“Did you mean…?”, “Constrain to…”, “Undo last interpretation”) as first-class actions, not error states (c.f. \cite{suchmanHumanMachineReconfigurationsPlans2007}).
\item \textit{Grounding / common ground}. Progress depends on establishing mutual belief about what has been done and what is intended \cite{clarkGroundingCommunication1991}. Persistent interaction history, lightweight provenance, and explanations of reconfiguration support grounding.
\item \textit{Indexicality \& situatedness}. Meanings are occasioned by local context (what was just drawn, selected, or said). The system should treat recent activity as indexical context for interpretation, not merely parse decontextualized input \cite{suchmanHumanMachineReconfigurationsPlans2007}.
\end{itemize}

This makes the metaphor useful: the aim is not to “chat”, but to sustain a sequential exchange where each move both responds to the prior state and sets up what comes next. The temporal logic of such exchanges also legitimizes pause, ambiguity, and revision: clarifications, restatements, and deferrals are ordinary repair moves rather than failures.

Crucially, “conversation” here is multimodal. Turns can be text, gesture, sketch, or selection. What matters is the sequential organization (turn→next-turn relevance, repair, grounding), not the linguistic channel. Where the desktop metaphor emphasizes what the system can do (a priori repertoire), the conversation metaphor emphasizes how participants come to know what to do together over turns. In hermeneutic terms, interaction becomes a fusion of horizons achieved sequentially rather than by selecting from fixed representations \cite{gadamerTruthMethod2013}. We do not ascribe mental states to LLMs; “conversation” names the interactional organization, not understanding or intention.

This notion of sequential organization builds upon extensive conversation analysis and conversational user interface (CUI) research on progressivity, repair, and co-construction in conversational interfaces \cite{porcheronVoiceInterfacesEveryday2018, reevesThisNotWhat2018, fischerProgressivityVoiceInterface2019, ashktorabResilientChatbotsRepair2019}. However unlike conversational interfaces, where conversation is the primary mode of interaction, LLM-mediated computing integrates conversational structure into a broader multimodal exchange. This integration makes the conversation not an interface modality but a computational mechanism, where each exchange can alter the substrate of computing itself. Conversation configures computation, rather than merely mediating access to fixed functions.

\subsection{Reflective Conversation}

The conversational structure outlined above helps us describe how interaction unfolds over turns, but it does not yet specify the quality of engagement we are interested in. To clarify the kind of conversation, and to avoid anthropomorphization, we turn to a related but distinct concept: \textit{reflective conversation}. In his analysis of \textit{designing as reflective conversation} \cite{schonDesigningReflectiveConversation1992}, Schön shows how designers engage in a dialogue with their materials through cycles of seeing–moving–seeing. A designer makes a \textit{move} – a transformation expressing an intention – then attends to how the situation “talks back” through its consequences, both intended and unintended. This backtalk may affirm the move or reveal a mismatch, prompting the designer to \textit{reframe} the situation and set a new problem. Through this ongoing exchange, a designer and situation co-constitute a provisional “design world” that evolves with each turn. For Schön, reflective conversation is explicitly metaphorical, making it apt for our search for a new metaphor in LLM-mediated computing.

We propose that LLM-mediated computing introduces a similar structure of engagement for everyday users. An intent is a \textit{move}, an expressive act. The LLM’s reconfiguration of the computer is a form of \textit{backtalk} – not intentional, but consequential – revealing interpretations, affordances, or misalignments. The user evaluates this response, \textit{reframes} their intent, and tries again. Through this back-and-forth, the system’s capabilities and the user’s purposes are co-articulated.

This reflective conversation is not smooth or linear. As in Schön’s account, surprises and misunderstandings are productive. The LLM might misinterpret intent, offer an unexpected alternative, or surface a new possibility. This exchange is structurally asymmetric: the human articulates intention, while the LLM-computer pair responds interpretively rather than intentionally. Its “reflection” is algorithmic, not cognitive. Yet it is precisely this interpretive gap that sustains reflection, inviting the user to rearticulate and redirect. The system becomes a partner in thought, not because it shares understanding, but because it supports exploratory articulation.

These dynamics resonate with prior studies of conversational interaction in HCI and CUI, which describe how sequential organization, progressivity, and repair structure human–system exchanges \cite{porcheronVoiceInterfacesEveryday2018, reevesThisNotWhat2018, fischerProgressivityVoiceInterface2019}. However, unlike these accounts, which analyze conversationality within fixed interactional modalities (e.g., voice or chat), reflective conversation concerns a more fundamental substrate: the computer itself becomes reconfigurable through these exchanges. In this sense, the sequential and negotiated structure described in prior work forms the interactional grammar through which LLM-mediated computing unfolds as reflective, material engagement.

Our use of reflective conversation extends beyond Schön’s original account: from professional designers’ interaction with material to the everyday use of computational media. Unlike Reflective Design \cite{sengersReflectiveDesign2005a}, which sought to provoke reflection upon technology, reflection here is intrinsic to the ongoing relation with a generative system.

We offer reflective conversation as a new metaphor for computing. It serves as both an analytic and prescriptive concept. Analytically, it describes how meaning and capability emerge sequentially through interaction. Prescriptively, it serves as a heuristic for designing systems that support such interpretive reciprocity. To make this metaphor concrete, we next illustrate how reflective conversation unfolds as a sequence of moves and backtalk in the garden vignette.

\subsection{Illustrative Dynamics}

Returning to the garden vignette: as the user sketches paths and flowerbeds, the system overlays interpretations. It suggests plant types, adds a seasonal slider, and introduces simulation features. Each gesture becomes a move, and the system’s responses serve as backtalk: the user proposes form, the system offers possibilities, the user revises or accepts. The computer itself shifts roles, such as design assistant, simulator, communication channel, depending on the trajectory of the exchange.

These iterative exchanges can be understood as moments of design-in-use or appropriation \cite{carrollCompletingDesignUse2004, salovaaraAcceptanceAppropriationDesignOriented2009, krischkowskyCaptologyTechnologyAppropriation2016}. In the garden vignette, each sketch, slider adjustment, or accepted suggestion does not merely adapt a pre-given tool, but alters the very configuration of what the system can propose next. In contrast to traditional accounts, each move in LLM-mediated computing actively reconfigures what the computer is and can do. The interface is not merely appropriated but continuously redesigned in interaction. This turns appropriation into a dynamic loop of reciprocal adaptation, where both human and system participate in configuring the evolving computational environment.

This illustration shows successful reflective conversation, but also hints at challenges. Misinterpretations may arise: the system might suggest inappropriate plants, misread gestures, or misalign with user intent. Repair mechanisms become crucial: the user may need to clarify, constrain, or redirect the system’s interpretations. Grounding is essential: both user and system must maintain a shared understanding of what has been established in the evolving garden design and what is intended next. These dynamics underscore that reflective conversation is an interactional structure requiring attentiveness, responsiveness, and mutual adjustment.

In the next section, we develop how this sequential reciprocity becomes ontological: how user and system co-disclose the computer through their interaction.

\section{A Postphenomenological Analysis of LLM-Mediated Relations}

In this section we turn to postphenomenology to analyze how LLM-mediated computing reconfigures human-computer relations and leads to what we call co-disclosure. Reflective conversation helps describe how this mode of computing unfolds in practice, but to understand its deeper significance we draw on postphenomenology \cite{ihdeTechnologyLifeworldGarden1990}, which analyzes how technologies mediate human experience and shape how the world is perceived, interpreted, and acted upon. Postphenomenology distinguishes relations such as embodiment, hermeneutic, alterity, and background, each of which discloses aspects of the world in specific ways.

In LLM-mediated computing, however, it is not only the world that is disclosed through technology. What is at stake is the disclosure of the computer itself. Each interaction configures what the computer is in the moment of use: a calculator, a simulator, a collaborator, or a messenger. This calls for extending postphenomenological analysis from interpreting how technologies mediate access to the world, to examining how the very substrate of computing is constituted in interaction. 


\subsection{Mapping and Extending Postphenomenological Relations}

Postphenomenology provides a vocabulary for describing how technologies mediate human–world relations. Ihde describes four recurring types \cite{ihdeTechnologyLifeworldGarden1990}: embodiment relations, where a technology extends the body (such as a cane or touchscreen); hermeneutic relations, where the world is presented through representations (such as a graph or thermometer); alterity relations, where a technology appears as a quasi-other (such as a robot or chatbot); and background relations, where technologies recede into the environment yet still structure experience (such as lighting or infrastructure). These categories help describe how technologies shape perception, interpretation, and action.

In LLM-mediated computing, these relations are not neatly separable. Embodiment, hermeneutic, and alterity dynamics overlap as users draw, type, or gesture while the system interprets and reconfigures the environment. More importantly, the object of mediation itself shifts. Traditionally, the computer mediates access to the world: a word processor mediates writing, a spreadsheet mediates calculation, a search engine mediates knowledge retrieval. In LLM-mediated computing, what is mediated is not only the world but the computer itself. The interface, capabilities, and even the identity of the system emerge in the interaction rather than being fixed in advance.

In our garden vignette, all four relations can be observed: the stylus embodies drawing, the seasonal slider is hermeneutic, the system’s suggestions appear as alterity, and the simulation engine operates in the background. Table~\ref{tab:relations} summarizes this mapping. Yet none of these relations fully capture the way the computer itself is constituted in use. To account for this, we add a fifth: co-disclosure.

\begin{table*}[ht]
\centering
\begin{tabular}{p{2.8cm}p{4.5cm}p{6.5cm}}
\toprule
\textbf{Relation} & \textbf{Structure (Human--Technology--World)} & \textbf{Garden vignette example} \\
\midrule
Embodiment & Human $\rightarrow$ (Technology) $\rightarrow$ World & The stylus and tablet withdraw; the user is immersed in shaping paths and flowerbeds. \\
\addlinespace
Hermeneutic & Human $\rightarrow$ Technology ($\rightarrow$ World) & The seasonal slider and annotations represent the garden’s future states; the user interprets these signs. \\
\addlinespace
Alterity & Human $\rightarrow$ Technology (as quasi-other) & The system suggests collaborators or water filtration, addressing the user as if in dialogue. \\
\addlinespace
Background & (Technology) $\rightarrow$ World / Human & Simulation runs quietly, animating ripples and seasonal overlays without explicit commands. \\
\addlinespace
\textbf{Co-disclosure (proposed)} & Human $\leftrightarrow$ LLM $\leftrightarrow$ Computer (constituted in interaction) & The very \textit{capabilities} of the system emerge through the sketching--suggestion loop: seasonal simulation, plant knowledge, and collaboration features come into being only as the conversation unfolds. \\
\bottomrule
\end{tabular}
\caption{Relations in Ihde’s typology and in the human--LLM--computer triad, illustrated through the garden vignette.}
\label{tab:relations}
\end{table*}

We describe this as \textit{co-disclosure relation}. In this relation, the human articulates intent, the LLM interprets and assembles functionality, and the computer reconfigures accordingly. Each turn discloses not just a new representation of the world, but a new configuration of the computer. Interpretation thus generates the very conditions for subsequent interpretation. The computer is not a stable mediator but an enacted participant whose identity is constituted in use. We therefore use co-disclosure both to name the overarching mode of LLM-mediated computing, and to identify a distinct relation within postphenomenology that supplements Ihde’s four, accepting this dual role because the same dynamic operates at both the systemic and relational levels.

This shift sets the stage for the dynamics we analyze in the following sections: feedback loops of observation and reconfiguration, the unfolding of mediation in situated examples, and the generative multistability through which the computer becomes different things in different moments.

\subsection{Dynamic Mediation and Observational Feedback Loops}

Using the metaphor of reflective conversation, we see that interaction in LLM-mediated computing is not a linear chain of input and output. It unfolds as a temporally sequenced process of mutual observation. The human expresses intent through words, gestures, or other actions. The LLM interprets these as signals, generates code or interface adjustments, and the computer reconfigures accordingly. The human then observes these reconfigurations, interprets them as meaningful, and adapts their activity.


Recall the generative feedback loop introduced earlier: human acts, LLM observes and interprets, computer reconfigures, human engages with new structures. Each cycle is an observational feedback loop where both human and system continuously adapt to each other. The human’s intentions evolve in response to the system’s reconfigurations, while the LLM’s interpretations are shaped by the unfolding context of use.

These recursive exchanges constitute what we call dynamic mediation: an unfolding process in which perception, interpretation, and reconfiguration feed into one another over time.

From a postphenomenological perspective, we see how we enter different relations with the technology at different stages in this loop:
\begin{itemize}
    \item \textit{Embodiment relations:} the human acts through the computer, which extends their expressive capacity via natural language or other inputs.
    \item \textit{Hermeneutic relations:} the computer’s reconfigurations, such as a text field becoming a “Send to Stephanie” button, are interpreted as signs that confirm or misread intent.
    \item \textit{Alterity relations:} the LLM sometimes appears as a quasi-other, anticipating needs or offering suggestions the human did not specify.
    \item \textit{Background relations:} even cursor movements, pauses, or deletions become tacit signals, drawn into mediation as background cues of intent or uncertainty.
\end{itemize}

Taken together, we see how the familiar Heideggerian distinction between ready-to-hand and present-at-hand begins to break down. Since the computer is no longer a stable tool, but a tool that is constructed while in use, there is nothing ready-to-hand. At the same time, it is \emph{in use} and thus not exactly present-at-hand. The computer is therefore better thought of as a medium than a tool, in line with our postphenomenological account of LLM-mediated computing. This breakdown makes visible why co-disclosure is needed: the computer cannot be captured as either tool or object, but as something \textit{constituted in use}.




As seen in Table~\ref{tab:relations}, each of Ihde’s relations can be observed in our garden vignette. Yet what is most revealing is not their separateness, but their simultaneity and fluidity. Embodiment, hermeneutic, alterity, and background relations do not occur in isolation but overlap and shift as the interaction unfolds. The stylus extends the hand in one moment, the seasonal slider calls for interpretive reading in the next, while system suggestions appear as an alterity presence, and background simulations quietly animate the scene. 

This layering of relations makes visible how mediation in LLM-mediated computing is not static but dynamic. Relations oscillate rapidly, sometimes even within the same gesture: drawing a pond embodies expression, triggers hermeneutic reinterpretations, and elicits alterity suggestions, all while background processes ripple the water. At the center of these shifts is the LLM, which interprets the human’s expressions and reconfigures the computer environment in response. What is disclosed here is not only the garden as a projected world, but the computer itself as a mutable environment, enacted turn by turn through this interpretive loop. Each cycle reconfigures both what the human is working on, and the environment in which that work is taking place.

\subsection{Multistability and Co-Disclosure}

These layered and shifting mediations point toward a deeper condition: the generative multistability of the computer itself. In postphenomenology, multistability describes how technologies do not have one fixed meaning or function, but can be taken up differently depending on context \cite{ihdeTechnologyLifeworldGarden1990,deboerExplainingMultistabilityPostphenomenology2021}. A hammer can be a tool or a weapon; a smartphone can be a phone, a camera, or a flashlight. Technologies become what they are in relation to human activity.

In LLM-mediated computing, this multistability is no longer a set of latent possibilities awaiting appropriation. Instead, stabilities are generated dynamically in the course of interaction. A computer becomes a calculator, a simulator, or a collaborative editor not because these functions pre-exist and are selected, but because they are assembled on the fly as the LLM interprets unfolding activity. Multistability becomes generative, arising sequentially as the relation develops. We call this dynamic \textit{co-disclosure}. The computer’s role is not pre-given but co-constituted through the interpretive loop of human expression and LLM continuation. Each stability emerges provisionally in relation and may dissolve or transform as interaction continues. The computer is not simply multistable in the sense of having many possible uses; it is enacted as different things in sequence, disclosed with the human in use.

A parallel can be found in Schön’s account of designing as worldmaking. Designers, he argues, construct the ontology of the design situation through seeing–moving–seeing – naming features, discovering unintended consequences, and transforming the situation as they act \cite{schonDesigningReflectiveConversation1992}. Co-disclosure extends this logic from professional design to everyday computing: here the “material” is interpretive, generating reconfigurations that act as backtalk and participate in reframing what the computer becomes in each moment.

This rethinking of multistability has two key implications. First, it reframes the ontology of the computer. The computer is not a fixed substrate onto which predefined functions are projected, but an emergent configuration that becomes what the relation calls forth. Second, it repositions the human–technology relation. The user is not merely selecting from existing stabilities but participating in their creation. The system is not merely responsive, but constitutive of the environment in which action becomes possible.

In this sense, co-disclosure extends postphenomenology. It suggests that technological mediation does not only shape how the world is disclosed, but also how the identity of the technology itself comes into being. LLM-mediated computing exemplifies this condition: a mode of computing, and a relation, in which world and computer are disclosed together, recursively and provisionally, in interaction.

\section{Discussion}

LLM-mediated computing, as we have argued, reframes the computer as something co-disclosed in interaction rather than given in advance. This shift has consequences that reach beyond interface design or user experience. It calls for a reconsideration of what it means to use a computer, how systems should be designed to support interpretive engagement, how evaluation can account for emergent trajectories, and how politics and ethics are embedded in the very conditions of disclosure. It also requires rethinking the conceptual resources – our theories, models, and metaphors – that guide how we understand and design these systems. In what follows, we discuss these implications across four dimensions: relational engagement, design, evaluation, and politics, before turning to how theory and metaphor can help orient design practice within this new paradigm.

\subsection{Reframing “Use” as Relational Engagement}

The concept of “use” has long structured human-computer interaction. In classical HCI, to use a computer is to operate a machine toward a predefined goal: opening a file, executing a command, completing a task. Even when later paradigms emphasized situated action \cite{suchmanHumanMachineReconfigurationsPlans2007} or distributed cognition \cite{hollanDistributedCognitionNew2000}, the computer itself typically remained a stable substrate that users acted upon.

LLM-mediated computing unsettles this model. The system is no longer just a set of capabilities to be accessed; it becomes an evolving presence whose behavior is shaped through interaction. The “computer” itself, as in the thing we act with, think through, and respond to, emerges dynamically as a function of engagement. What we call the computer in this context is not a fixed tool, but a configuration that is co-constructed in real time by user intent and system response.

This shift moves us from a model of tool use to one of relational engagement. Drawing on relational theories of technology (e.g., Suchman, Verbeek, Ihde \cite{ihdeTechnologyLifeworldGarden1990,suchmanHumanMachineReconfigurationsPlans2007,verbeekWhatThingsPhilosophical2005}), we can begin to see interaction not as a linear chain of actions, but as a temporal and interpretive relationship. The system responds to the user, but also observes and adapts to how the user is responding to it. Each input is shaped by what came before, and anticipates what might follow. “Use” becomes less like command execution, and more like participating in an evolving interactional trajectory.

Reframing use as relational engagement has several implications. First, it positions the human not only as operator but as participant and interpreter, responsible for making sense of the system’s responses and adjusting their own actions accordingly. Second, it positions the system not as a passive instrument but as an active presence, proposing possibilities and reshaping the field of action. Finally, it redefines the object of design: what is at stake is not a fixed set of functions or interfaces, but the quality of the interactional relation itself. How it supports continuity, reflection, and mutual intelligibility over time.
%
To “use” a computer in this sense is to participate in a reflective conversation where the system’s identity and the user’s trajectory are shaped together.

This orientation resonates with the broader “entanglement turn” in HCI, where researchers have emphasized that humans and technologies are inseparably bound in sociomaterial assemblages rather than divided into users and tools \cite{frauenbergerEntanglementHCINext2019}. Our account complements this view but extends it in a different direction. With LLM-mediated computing, computing is not only recognized as relational through ontological entanglement, but becomes relational by default. The very capabilities of the computer, what it can do and what it can be used for at any given moment, are not given in advance but co-disclosed through ongoing interaction. This has important consequences for design: if computing is relational by default, then what must be designed is not only interfaces or functions, but the very quality of the relation through which the computer comes into being.

\subsection{Designing for Reflective Conversation}


Designing for LLM-mediated computing means acknowledging that meaning is provisional, responses are open to revision, and the computer itself is constituted through the exchange. This places new demands on design:

\begin{itemize}
\item \textit{Legibility of mediation.} If the system introduces features, adapts tools, or reconfigures the interface, users need to be able to see and understand how this happened. Without such legibility, mediation risks becoming opaque, leading to misplaced trust or confusion. Design must therefore surface the system’s role in shaping interaction, through explanations, interaction histories, or contextual cues, building on work in transparency and human-centered explainability \cite{liaoHumanCenteredExplainableAI2022}.
\item \textit{Support for tentativeness and repair.} Just as in conversation, misunderstandings are expected and productive. Interfaces should not reject vague or partial inputs, but scaffold them toward clarity. Repair should be treated as a natural part of interaction rather than a failure. In conversational terms, this is the design of repair moves: ways for users to clarify, restate, or undo, rather than treating breakdowns as errors. This aligns with CA-informed analyses of breakdowns in CUIs, which show how repair practices shape interaction \cite{ashktorabResilientChatbotsRepair2019,porcheronVoiceInterfacesEveryday2018}.
\item \textit{Visible interaction history}. Reflective conversation requires memory. Users must be able to revisit, revise, and reinterpret earlier exchanges, not just see a final output. Design can support this through traceable interaction histories that foreground how meaning and functionality have evolved, echoing work on interaction traces and data literacy in voice and AI-mediated systems \cite{pinsAlexaWeNeed2021}.
\item \textit{Multimodal responsiveness}. In practice, interaction is not only textual. Systems should interpret and respond across modalities, such as voice, sketch, or gesture, so that reflective conversation is not constrained to a single channel. This resonates with recent work on multimodal and reflective AI systems that support real-time sensemaking across modalities \cite{wenPromotingRealTimeReflection2025,glinkaCriticalReflectiveHumanAICollaboration2023}.
\item \textit{Temporal flexibility}. Reflection takes time. Systems should allow for pauses, digressions, and deferred continuation, supporting rhythms beyond immediacy. Temporal openness ensures that interaction can unfold as an extended interpretive process rather than a rapid-response exchange.
\end{itemize}

Designing for reflective conversation means designing for a relation in which users can see how mediation is unfolding, respond to it critically, and carry responsibility for how the computer is constituted in use. The principles above extend prior work on transparency, repair, history, and multimodality by treating them not as interface features but as conditions for a shared interpretive process. Reflective conversation does not aim to make computers more human-like; rather, it acknowledges interaction as a relational and situated activity and builds systems that can participate in that activity meaningfully, even if incompletely. This opens a distinct design space in which the goal is to sustain an evolving dialogue through which functionality and meaning are continually reconfigured.

Human-centered design of computing gets closer to human-centered AI \cite{shneidermanHumanCenteredAI2022} than to traditional usability measures. These design demands align closely with ongoing work in Human--AI Interaction and Explainable AI. Amershi et~al.\ \cite{amershiGuidelinesHumanAIInteraction2019} outline guidelines that emphasise the importance of making system capabilities visible, supporting correction and adaptation, and maintaining continuity over time. Similarly, Liao et~al.\ \cite{liaoHumanCenteredExplainableAI2022} highlight the need for situated, user-adaptive explainability that evolves with interaction.  Research in Human-Centered AI further underscores how interpretive negotiation and mutual adjustment shape collaboration between humans and AI systems \cite{weiszDesignPrinciplesGenerative2024}. Reflective conversation aligns with these perspectives by viewing legibility, repair, and accountability as central to sustaining a shared understanding through interaction, grounding these principles in the sequential and interpretive dynamics of human--AI exchange.



\subsection{Rethinking Evaluation}

If interaction with LLM-mediated systems is best understood as reflective conversation, then existing evaluation approaches in HCI fall short. Measures such as efficiency, error rates, or task completion presume predefined goals and stable system functions. In contrast, LLM-mediated computing unfolds as a trajectory and evaluation in this paradigm must therefore shift focus:

\begin{itemize}
\item \textit{Trajectory over outcomes}. Success lies not only in task completion but in how meaning and capability develop across turns. Evaluation should attend to the coherence and richness of the interactional path.
\item \textit{Responsiveness to ambiguity}. A key quality of reflective conversation is handling partial or uncertain input. Systems should be assessed on whether they scaffold vague expressions into workable directions rather than rejecting them outright.
\item \textit{Repair and reflection}. Breakdowns are not failures but moments for redirection and learning. Evaluation should examine how systems support repair, revision, and reflection-in-action.
\item \textit{Epistemic inclusivity}. LLM-mediated computing privileges some forms of knowing (e.g. articulate, linguistic expression) over others (e.g. tacit or embodied practice). Evaluation must ask whose knowledge practices are supported, and whose are marginalized.
\end{itemize}

These shifts point toward more qualitative and relational methods: examining interaction trajectories, studying repair and improvisation, and attending to the interpretive conditions of engagement. To evaluate reflective conversation is to evaluate a relation, not only an artifact or interaction.

This orientation also resonates with Gaver’s argument that ambiguity can be a resource for design rather than a problem to be eliminated \cite{gaverAmbiguityResourceDesign2003}. Where Gaver emphasizes deliberately crafted ambiguity to invite reflection, LLM-mediated computing introduces ambiguity as an ordinary condition of interaction: interpretations are provisional, system responses are open to revision, and meaning develops across turns. Evaluation therefore cannot treat ambiguity as error, but must examine how systems support users in navigating, clarifying, and productively engaging with it throughout the interaction trajectory.

To make these principles more actionable, we outline two illustrative methods. We suggest two methods of evaluation, not as an exhaustive list but as a starting point for this paradigm. \textit{Trajectory Analysis} focuses on the arc of interaction and examines how intent, interpretation, and reconfiguration unfold across turns, with attention to continuity, coherence, and shifts in meaning. Researchers can code sequences for how user actions project system responses, how the system’s proposals extend or constrain possibilities, and how users subsequently adapt. \textit{Repair Audit} evaluates how systems support clarification, revision, and recovery, assessing whether repair is visible, lightweight, and productive rather than disruptive or hidden.

\subsection{The Politics of Mediation}

We now turn to politics. By politics we mean the distribution of power, values, and governance embedded in the infrastructures that shape interaction \cite{bowkerSortingThingsOut2000, suchmanHumanMachineReconfigurationsPlans2007}. Postphenomenology reminds us that mediation is never neutral: technologies always co-shape how the world is disclosed \cite{ihdeTechnologyLifeworldGarden1990,verbeekWhatThingsPhilosophical2005}. In LLM-mediated computing, this non-neutrality is inflected by prior decisions about training data, fine-tuning, safety filters, and interface design. Each reconfiguration of the environment, whether it involves which tools appear, which interpretations are privileged, or which possibilities are never proposed, is conditioned by such decisions. These infrastructural conditions are rarely visible to the user, yet they structure what the “computer” can become in practice \cite{molOntologicalPoliticsWord1999,agreSurveillanceCaptureTwo1994}. As Agre argues, such infrastructures always embody a “grammar of action” \cite{agreSurveillanceCaptureTwo1994}: implicit models of what forms of activity can be recognized, classified, and acted upon. While application-centered computing makes this grammar explicit, LLM-mediated computing does not remove it but renders it implicit. The constraints that shape what the system can interpret or propose are now embedded in training data and alignment processes, making the grammar of action harder to see yet no less powerful in structuring interaction.

This raises several political questions \cite{birhaneAlgorithmicInjusticeRelational2021}. Whose conversational norms and languages are recognized? Whose values are encoded in the system’s default responses? Which practices of knowing are supported, and which are rendered invisible? What appears to be a responsive partner in reflective conversation is also an artifact of upstream design choices and institutional governance.

These concerns resonate with broader critical debates on the politics of AI. Scholars have shown how large language models are never neutral: their training data, fine-tuning, and deployment embed cultural values, exclusions, and asymmetries of power \cite{crawfordAtlasAIPower2021, benderDangersStochasticParrots2021,nobleAlgorithmsOppressionHow2018,benjaminRaceTechnologyAbolitionist2019}. Coeckelbergh \cite{coeckelberghAIEthics2020} argues that ethics in AI must be understood as inseparable from these political conditions, since questions of responsibility, justice, and democracy are already configured in the infrastructures that make AI possible. If, as we suggest, LLMs come to shape and form what the “computer” is in practice, then these ethical and political debates do not sit alongside LLM-mediated computing but are integral to it. Indeed, the stakes may be even higher: when mediation itself is constituted by LLMs, the politics of AI becomes the politics of computing. These debates make clear that mediation is always already entangled with political economy, culture, and governance, concerns that must be taken seriously within HCI’s conceptualizations of LLM-mediated computing.

The metaphor of conversation can obscure these dynamics. It presents interaction as mutual, open-ended, and shared. In reality, the exchange is asymmetrical. While the user adapts to the system’s proposals, the range of proposals is delimited by corporate, technical, and cultural infrastructures. If a certain possibility is never offered, it may not even be imaginable to the user.

For HCI, this means that designing for reflective conversation also entails designing for contestability and accountability. Users must be able to question why the system responded in a certain way, what alternatives were excluded, and whose priorities shaped those exclusions. Without such reflexive capacities, the computer disclosed in interaction risks naturalizing hidden biases and asymmetries.
To engage critically with LLM-mediated computing, HCI must therefore attend not only to the relational dynamics of reflective conversation, but also to the infrastructures of power that condition what can be said, done, and imagined within it.

\subsection{Theory, Metaphor, and Reflexive Design}

Designing for LLM-mediated computing requires conceptual scaffolding. Because system behavior is open-ended, emergent, and interpretive, designers cannot rely on stable functions or predefined affordances; they must instead draw on theory to make sense of how interaction unfolds. Phenomenology and postphenomenology help articulate how technologies mediate perception and action \cite{ihdeTechnologyLifeworldGarden1990,verbeekWhatThingsPhilosophical2005}, hermeneutics foregrounds the interpretive character of engagement \cite{gadamerTruthMethod2013}, and Schön’s idea of reflection-in-action offers a structure for understanding how users and systems co-develop trajectories of meaning and capability \cite{schonReflectivePractitionerHow2017}. Conversation analysis likewise provides a vocabulary for the sequential organization of interaction \cite{silvermanHarveySacksSocial1998}. These theoretical resources make LLM-mediated computing intelligible and actionable by illuminating the kinds of relations, expectations, and interpretive moves that unfold in practice.

At the same time, the concepts and metaphors we adopt do more than describe systems – they shape how those systems are designed and imagined \cite{ericksonWorkingInterfaceMetaphors1990,lakoffMetaphorsWeLive2003}. Our proposal of reflective conversation is therefore both a design resource and a generative fiction: it highlights reciprocity, temporality, and negotiation, but also risks suggesting more symmetry or understanding than LLMs in fact possess. As with the desktop metaphor, which eventually hardened into infrastructure, any new metaphor carries the possibility of sedimenting into unexamined assumptions that privilege certain cultures of interaction while marginalizing others.

For this reason, theory must be used reflexively. Metaphors such as reflective conversation should not be taken as literal descriptions or as normative templates, but as provisional lenses that illuminate some aspects of LLM-mediated computing while obscuring others. Their value lies in helping designers see patterns of mediation, repair, and co-disclosure that would otherwise remain implicit, while maintaining awareness of the political and cultural commitments such framings entail. Used in this way, theory and metaphor become tools for orienting design without constraining it, enabling systems that support interpretive engagement while remaining open to alternative ways of imagining what the computer can become.

\section{Conclusion}

In this paper we have argued that LLM-mediated computing invites a shift from the desktop metaphor and application-centered computing toward computing as \textit{reflective conversation}. In this paradigm, the computer is not a fixed tool but an emergent presence, co-disclosed in use through the interplay of user intent and LLM responsiveness. What the computer can do at any moment is not predetermined but arises relationally through ongoing interaction.


Together, our three contributions (C1-C3) reframe computing not as the operation of predefined applications but as participation in an unfolding relation where the computer’s very capabilities emerge provisionally in interaction. They also ground practical implications: for \emph{design} (legibility of mediation, repair, history, temporal flexibility), for \emph{evaluation} (trajectories, responsiveness to ambiguity, support for repair, epistemic inclusivity), and for \emph{politics and ethics} (contestability and accountability when mediation is conditioned by data, fine-tuning, and governance).

Like all conceptual work, our account is necessarily partial. We have emphasized philosophical analysis and speculative examples, but empirical studies are needed to explore how users actually inhabit reflective conversation in practice. Future research could examine how co-disclosure plays out in real-world systems, how literacy for this mode of computing can be cultivated, and how design can support accountability without foreclosing openness.

The future of HCI may not lie in faster or smarter machines, but in reclaiming computing as an activity of shared interpretation and co-disclosure.

\bibliographystyle{ACM-Reference-Format}
\bibliography{better}

\end{document}